# LDPC for QKD Reconciliation

Alan Mink and Anastase Nakassis

*Abstract*—We present the Low Density Parity Check (LDPC) forward error correction algorithm adapted for the Quantum Key Distribution (QKD) protocol in a form readily applied by developers. A sparse parity check matrix is required for the LDPC algorithm and we suggest using some that have been defined by the IEEE and ETSI standards organizations for use in various communication protocols. We evaluate the QKD performance of these various parity check matrices as a function of the quantum bit error rate. We also discuss the computational precision required for this LPDC algorithm. As QKD evolves towards deployment, complete algorithm descriptions and performance analysis, as we present, will be required.

*Index Terms*— Quantum key distribution, QKD, Pseudo code, LDPC.

## I. INTRODUCTION

THE Quantum Key Distribution (QKD) protocol [1] uses an unsecured quantum channel and an unsecured, but integrity protected classical channel to establish a shared secret between two parties, Alice and Bob, at the two ends of the channels. There are four stages to the QKD protocol, see Fig. 1. Stage 1 is where the transmission of randomly encoded single photons occurs over the lossy quantum channel and the photons' value is measured. Stage 2 is where Alice and Bob exchange information over the classical channel to "sift" their bit sequences to achieve a common sequence to work with, but that sequence may have errors. Stage 3 is where Alice and Bob exchange information over the classical channel to reconcile, correct errors, between their bit sequences without exposing the value of their bits. Stage 4 is where Alice and Bob privacy amplify their now identical bit sequences through the application of a hash function that does not require any communication, yielding a shared secret between Alice and Bob.

We are focusing on the third stage of this protocol, the Reconciliation stage, where a rough secret (a sequence of bits called the sifted bits) has been established at both Alice and Bob but there are some errors in Bob's sequences of bits, although their number and order are the same as Alice's bits. The error rate within these sequences is called the quantum bit error rate (QBER) and must be less than 11% to provide the desired information theoretical security. High levels, as well as significant changes, in the QBER indicate potential eavesdropping. A Reconciliation algorithm to correct Bob's bits must do so indirectly, without exposing secret bit values. Any Reconciliation algorithm will, however, indirectly expose some information about those secret bit values. Unlike conventional communications, the QKD protocol uses two different channels; a lossy quantum channel for the original photon stream of the first protocol stage and a reliable classical channel for the other protocol stages.

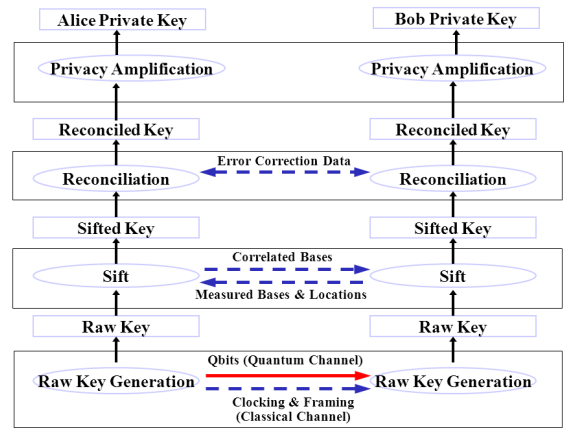

Figure 1. The four stages of the QKD protocol.

The Low Density Parity Check (LDPC) forward error correction algorithm was initially proposed by Gallager [5]. It became popular in the early 2000s for data communications and was first applied to the QKD protocol by BBN [4]. The main benefits of using LDPC for QKD are that only a single round trip communication is required and the amount of information that might be exposed to an eavesdropper is more easily computed compared to the Cascade [2] error correction algorithm, initially used for QKD, that requires a number of round trip communications. On the other hand, from our implementations, Cascade requires about 1 to 2 bytes of memory per bit of data to be corrected, whereas LDPC requires about 20 to 30 bytes of memory – an order of magnitude larger. This only accounts for required memory bits, not the larger amount actually allocated due to the fragmentation of minimum allocation units. For example, 18 bit values on a PC would be allocated in units of 32-bits,

Alan Mink is with the National Institute of Standards and Technology, Gaithersburg, MD 20899 USA (phone: 301-975-5681; fax: 301-975-6238; e-mail: amink@nist.gov).

Anastase Nakassis is with the National Institute of Standards and Technology, Gaithersburg, MD 20899 USA (e-mail: anakassis@nist.gov).

~75% additional overhead, and in FPGA hardware, 1000 10-bit entries would be allocated as a block of 1024 by 18-bits, ~85% additional overhead. For our FPGA implementations [11], Cascade's processing rate, ~5 Mbits/s, is also faster than LDPC's, between ~2 Mbits/s (for large dataset sizes, ~50K) and 4 Mbits/s (for small dataset sizes, ~2K), but the communication delays are negligible for these measurements. As the distance between Alice and Bob increases, communication delays will have a greater impact on Cascade's performance and thus favor LDPC. But even at 100 km, Cascade's performance only drops to ~3 Mbits/s because of its large computational granularity while LDPC's performance drops to ~1.5 Mbits/s for the small dataset size but stays approximately the same at about 2 Mbits/s for the large dataset size. These are per-thread execution rates and would linearly scale with parallel instantiations, but in hardware the number of threads is limited by available memory.

We discuss how the QKD environment differs from a classical communication channel environment and its impact on error correction. Nakassis [12] has developed efficient versions of the LDPC algorithms that can use fast table lookups rather than execute more time consuming mathematical functions. The trade-off is the accuracy required by these tables, which in turn translates to the size of those tables. We present a version of that algorithm in a form readily applied by developers. Furthermore, a major consideration in applying LDPC is how to build an appropriate parity check matrix. We propose adopting matrices already in use by some of the existing communication standards groups. Through an implementation of this LDPC algorithm, we evaluate the error correction performance of a number of these matrices as well as the precision of these lookup tables necessary to drive them and some heuristics that enhance their performance. As QKD evolves towards deployment, complete algorithm descriptions and performance analysis, as we present here, will be required.

## II. CLASSICAL VS. QKD ENVIRONMENT FOR LDPC

In the classical communication environment a sequence of data bits, $\mathbf{x}=\{x_0, x_1, \ldots, x_{n-1}\}$, is sent over an error prone channel resulting in a received sequence of data bits, $\mathbf{y}=\{y_0, y_1, \ldots, y_{n-1}\}$. Any differences between $\mathbf{x}$ and $\mathbf{y}$ are errors. LDPC applied to such an environment adds another sequence of parity bits, $\mathbf{s}=\{s_0, s_1, \ldots, s_{m-1}\}$, that is also sent over the same error prone channel. It is used to determine errors and help correct them. The transmission is the k-bit (k=n+m) sequence, $\mathbf{I}=\{x_0, x_1, \ldots, x_{n-1}, s_0, s_1, \ldots, s_{m-1}\}$, that is the concatenation of $\mathbf{x}$ and $\mathbf{s}$. The received sequence is $\mathbf{J}=\{y_0, y_1, \ldots, y_{n-1}, t_0, t_1, \ldots, t_{m-1}\}$, where $\{t_0, t_1, \ldots, t_{m-1}\}$ is the sequence of the received parity bits. Any differences between $\mathbf{I}$ and $\mathbf{J}$ are errors. Sender and receiver use the same predefined m-row by k-column sparse parity check matrix, $\mathbf{M}$, such that the syndrome is

$$[\mathbf{0}] = [\mathbf{M}] \text{ x } [\mathbf{I}]^T = [\mathbf{M_1} \mathbf{M_2}] \text{ x } [\mathbf{x} \text{ } \mathbf{s}]^T$$

where $[\mathbf{I}]^T$ is the transpose of $[\mathbf{I}]$, all arithmetic is modulo base 2 (i.e., additions are the same as XORs), $[\mathbf{0}]=\{0_0, 0_1, \ldots, 0_{m-1}\}$ (i.e., all zero elements), $\mathbf{M_2}$ is an m x m matrix that allows easy computation of the parity bits so that the matrix multiplication results in $[\mathbf{0}]$ (thus doesn't need to be sent to the receiver) and $\mathbf{M_1}$ is an m-row by n-column parity check matrix. The matrix $\mathbf{M}$ doesn't change for each new data sequence, but the parity bits, $\mathbf{s}$, must be recomputed for each new data sequence.

When the receiver gets its copy of the parity bits, $\mathbf{t}$, the receiver performs a similar computation:

$$[\mathbf{0}] = [\mathbf{M}] \text{ x } [\mathbf{J}]^T = [\mathbf{M_1} \mathbf{M_2}] \text{ x } [\mathbf{y} \text{ } \mathbf{t}]^T$$

If the result isn't $[\mathbf{0}]$, then there are errors in $\mathbf{J}$, and the belief propagation part of the LDPC algorithm is iteratively applied to revise $\mathbf{J}$ until it converges (i.e., the results are $[\mathbf{0}]$) or the maximum number of iterations is reached.

In classical wireless communications, the expected average error rate is $\sim 10^{-3}$ or less. The success or failure of the error correction may or may not be sent back to the transmitter. If LDPC fails, then additional forward error correction codes as well as retransmission are always an option.

In a QKD environment a sequence of qbits is sent over an error prone quantum channel in stage one of the QKD protocol, but only a small number of those bits are received. In stage two of the QKD protocol, only appropriately measured qbits received by Bob are "sifted" and retained by both Bob and Alice. This results in a similar condition to classical communications, where Alice has a bit sequence $\mathbf{x}=\{x_0, x_1, \ldots, x_{n-1}\}$ and Bob has a bit sequence $\mathbf{y}=\{y_0, y_1, \ldots, y_{n-1}\}$. Any differences between $\mathbf{x}$ and $\mathbf{y}$ are errors. Because LDPC in QKD, see Fig. 2, uses a reliable classical channel there is no need for the extra parity bits, $\mathbf{s}$, used in the classical case, resulting in a more streamlined approach as follows:

$$[\mathbf{CS}] = [\mathbf{M_1}] \text{ x } [\mathbf{x}]^T$$

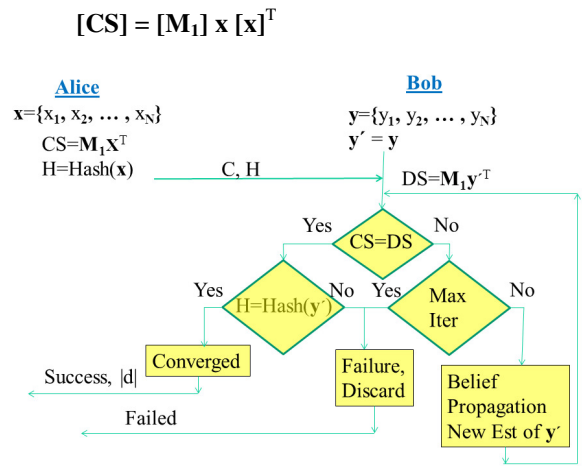

Figure 2. The QKD Reconciliation stage using LDCP.

But the resulting syndrome $[\mathbf{CS}]$, also called a checksum vector, is now a non-zero m-bit vector. So Alice sends $[\mathbf{CS}]$ to Bob over the reliable classical channel, which is received error free. Bob computes:

$$[DS] = [M_1] \text{ x } [y]^T$$

and compares the two. If $[CS] \neq [DS]$, then there are errors in **y**, and the belief propagation part of the LDPC algorithm is iteratively applied to correct **y** until it converges (i.e., the results are $[CS] = [DS]$) or the maximum number of iterations is reached. A hash signature is further used to verify, with high probability, that **y** has been corrected and the number of errors corrected is shared with Alice to update the QBER estimate.

The average QKD QBER is between $10^{-2}$ and $10^{-1}$, much larger than the classical communication cases. A QBER above 11% would result in discarding both **x** and **y** since security could no longer be relied upon. The success or failure of the error correction is always sent back to Alice. If LDPC fails, additional forward error correction codes are not an option due to concerns of exposing too much information about **x** and **y** and retransmission is not possible. Instead, both **x** and **y** are discarded.

### III. MATRICES AND TABLES

One problem in applying LDPC is building the matrix, $[M_1]$. The basic guidelines are that each row defines a single checksum, and usually contains somewhere between 5 and 20 "1"s while the rest of the elements of that row are "0"s. Each column represents a bit and indicates which checksums that bit participates in, and usually contains somewhere between 3 and 13 "1"s while the rest of the elements of that column are "0"s. Since the columns number in the thousands to tens of thousands and the number of rows is between 20% and 70% of the number of columns, we see that such matrices are indeed sparse. If the number of "1"s in each row is r and the number of "1"s in each column is c (r and c can be different), then the matrix is referred to as being regular. If r is different for different rows and/or c is different for different columns, then the matrix is referred to as irregular. Irregular matrices have been shown to provide better error correction performance than regular matrices [13]. As the error rate increases, the "1"s per row decrease, resulting in smaller checksums (i.e., fewer bits per checksum) but more of them (i.e., more rows).

A number of papers [9, 14] suggest a randomly generated matrix performs well. Others have noted that the occurrence of "cycles" in a matrix structure can affect the convergence of the algorithm. Although cycles cannot be eliminated in such matrices, their length can be maximized thus reducing their effects. One would also assume that properties of the channel errors could also have an effect on the selection of an optimum matrix. We have focused on matrices from standards groups that apply LDPC and have published their family of acceptable matrices, such as the IEEE 802.11n Std [15], IEEE 802.16e Std [16] and ETSI DVB Std [3]. These matrices have a systematic structure that allows a compact description for their generation. Alternatively we can follow construction guidelines presented by others [7, 10]. In a section below, using selected matrices from the above standards, we evaluate their error correction performance for QKD applications as a function of the QBER.

To compute the belief values in the belief propagation algorithm, we can do the floating point computations on demand or to save time we can map these values onto integers and prebuild a lookup table that would simplify the computations. This would result in faster execution times and would be attractive for hardware environments where arithmetic function support may be limited. We found a 1K size table of 10-bit entries (10-bit precision) adequate for many situations, but not all. A 2K size table of 11-bit entries (11-bit precision) yielded better results, but a 4K size table of 12-bit entries (12-bit precision), worked well for all our test cases. We define the following profiles, one for each of these different size tables:

Profile 1 yields a ~1K table of 10 bit entries: Na=175, Ma=1023, Nf=173, Mf=1013.

Profile 2 yields a ~2K table of 11 bit entries: Na=312, Ma=2009, Nf=313, Mf=2009.

Profile 3 yields a ~4K table f 12 bit entries: Na=555, Ma=3893, Nf=556, Mf=3896.

Our algorithm uses natural logarithms to further simplify the computations to additions and subtractions and requires two lookup tables, an "a2f" and an "f2a" table. They were constructed from the pseudo code in Fig 3, using constants defined in the profiles above.

The initial belief values needed for the algorithm can also be pre-computed and stored in a pair of small tables of a dozen entries each. They can be constructed from the following equations:

F_init = Nf * ln( (1-p)/p );  // initial F belief values
A_init = -Na * ln( (1-2*p) ); // initial A values where Na and Nf are taken from the profiles above and used to map these decimal values to the integer precision desired, p is the estimated QBER, $0.01 \leq \text{QBER} \leq 0.12$. We always round the QBER up to the next multiple of 0.01 (e.g., a QBER of 2.1% is rounded to 3%). As an example, using profile 1 we obtain the belief initialization tables (in hexadecimal notation) shown in Table 1.

```
for (i=1; i<=Ma; i++)          for (i=1; i<=Mf; i++)
{  a = exp(i/Na);              {  f = exp(i/Nf);   // power of e
   f = (1+a)/(1-a);              a = (f-1)/(f+1);
   z = ln(f);                    z = -ln(a);   // natural logarithm
   j = (int) (Nf*z+0.5);         j = (int) (Na*z+0.5); // integer map
   if (j<=0) j=1;                if (j<=0) j=1;  // keep in limits
   if (j>Mf) j=Mf;               if (j>Ma) j=Ma;
   a2f[i] = j;                   f2a[i] = j;     // load table value
}                              }
       (a)                              (b)
```

Figure 3. Pseudo code for the construction of the LDPC (a) "a2f" and (b) "f2a" lookup tables.

## IV. LDPC Belief Propagation Algorithm

A complete pseudo code description of our version of the LDPC algorithm, including belief propagation, is presented in Fig 5. It is fashioned from the logarithm-based algorithm presented by Nakassis [12]. We define the following information and data structures used in this algorithm:

| | |
|---|---|
| y[1:n] | - a list of Bob's original info bits |
| y1[1:n] | - a list of Bob's corrected info bits |
| c[1:m] | - a list of Alice's checksum values |
| b[1:m] | - a list of Bob's original checksum values |
| d[1:m] | - c XOR b, XOR of Alice & Bob's checksum values |
| cs_list[1:k] | - a list of checksum info (the sparse matrix info) as follows: |
| cs_list [i].bit# | - the i-th participating bit number |
| cs_list [i].LL_ptr | - a pointer to the associated LL_reg |
| LL_reg[1;k] | - a list of the belief registers |

where n is the number of info bits, m is the number of checksum values and k is the number of non-empty entries in the sparse matrix. If the matrix is regular (all checksums contain the same number of bits and each bit participates in the same number of checksums), then k = (bits/checksum * m). If the matrix is irregular (variable checksum lengths and/or bits participating in a variable number of checksums), then one must compute the value of k by summing the number of bits in each checksum. In addition to these data lists, we also need mappings showing the start and end of each checksum in the cs_list list as well as a mapping showing the start and end of each bit's belief register group in the LL_reg list, see Fig 4. The cs_list defines each checksum, the participating bit numbers and its associated belief register, and is ordered by checksum. The i-th entry of the cs_index list points to the start of the i-th checksum definition. The LL_reg list contains the belief values and is ordered by bit number. The i-th entry of the LL_index list points to the start of the i-th bit belief register group.

| | |
|---|---|
| cs_index[j] | - ptr to the start of the j-th chksum (j=1, … , m+1), where cs_index[m+1]=k+1 |
| LL_index[i] | - ptr to the start of the LL_reg group for the i-th info bit (i=1, … , n+1), where LL_index [n+1]=k+1 |

## V. Performance, Precision & Heuristics

Performance for LDPC is commonly presented in the form of "waterfall graphs" that plot correctable error rates vs. signal-to-noise ratio (SNR). This is mainly done because classical communication uses various encoding schemes that map differently to the bit error rate. For QKD performance, a more useful presentation is a table showing correctability vs. QBER. For our analysis we selected the largest matrices from the IEEE 802.11n Std [15] (z=81), IEEE 802.16e Std [16] (z=96), and ETSI DVB Std [3] (FECFRAME=64800). The matrices are summarized in Table 2, which shows their code rates, the number of information bits, the number of checksum bits and the expected limits of their correction based on the QBER (i.e., the first entry in the table is expected to successfully correct data that has up to ~2% QBER based purely on entropy calculations). For a rough estimate of correctability, we estimated that rate ratio should exceed the entropy, H, by about 30%, 1.3*H < (1-rate)/rate, where H = -(p*log₂p+(1-p)*log₂(1-p)), p=QBER and rate is the rate of the LDPC matrix. For each of our measurements we ran 1000 different samples. For each sample, we generated random bits for Alice from the standard C pseudo random number generator (PRNG) that uses a uniform distribution. To generate Bob's information bits, we randomly flipped some of Alice's bits, based on the QBER and using the same PRNG. Thus the actual error rate of any sample varies about a mean whose value is the QBER.

| a_init_list[1:12] = { | f_init_list[1:12] = { |
|---|---|
| 0x0004,  // 1% | 0x0316,  // 1% QBER |
| 0x0007,  // 2% | 0x02A1,  // 2% |
| 0x000B,  // 3% | 0x0259,  // 3% |
| 0x000F,  // 4% | 0x0226,  // 4% |
| 0x0012,  // 5% | 0x01FD,  // 5% |
| 0x0016,  // 6% | 0x01DC,  // 6% |
| 0x001A  // 7% | 0x01C0,  // 7% |
| 0x001F,  // 8% | 0x01A7,  // 8% |
| 0x0023,  // 9% | 0x0190,  // 9% |
| 0x0027,  // 10% | 0x017C,  // 10% |
| 0x002B,  // 11% | 0x016A,  // 11% |
| 0x0030  // 12% | 0x0159  // 12% |
| }; | }; |
| (a) | (b) |

Table 1. Example (using profile 1, 10-bit precision) of initial belief (a) "a" and (b) "f" value lookup tables.

To establish a baseline for the performance of these matrices, we used a version of the algorithm presented above in which we employed double precision floating point variables and arithmetic instead of our integer mappings and replaced the table lookups with direct calculations (i.e., we used code similar to the first 3 lines of the of the pseudo code in Fig. 3, rather that the "a2f" & "f2a" tables). This implementation, running on a PC in the C programming language, was about an order of magnitude slower than our lookup table version on the same PC also in C. The results are shown in Table 3. Zero failure entries in the table indicates that all correction attempts were successful for the 1000 samples used in these test, but it doesn't mean that it can correct all possible error sequences for the QBER. There are some sequences that aren't correctable by this technique, but they would be a small number. As the "waterfall" graphs in other LDPC papers show, the resultant error rate, after correction, decreases exponentially as the signal-to-noise ratio increases (QBER decreases) but doesn't go to zero.

As we scan Table 3 from left to right, we see that the failures go from low to high. So this format makes it easy to determine at what QBER a given matrix begins to fail and how different matrices of the same rate provide different correction performance. In these experiments, the ETSI matrices outperform the IEEE matrices. For example, the ETSI 5/6 rate matrix corrects up to a QBER of 2% with very low failure rate, while the IEEE 5/6 rate matrices has a low failure rate at a 1% QBER, but incurs about a 40% failure rate at a QBER of 2% . We believe this is due to the ETSI matrices having a larger dataset than the IEEE matrices, more than an order of magnitude larger. This allows more separation of information bits in checksums, what is referred to as larger cycles [8]. Also the ETSI matrices have a sharper failure drop-off (i.e., going from one QBER value to the next results in complete failure) whereas the IEEE matrices seem to drop-off more slowly, over a number of QBER values.

For our table lookup implementation, we investigated a number of table sizes. The prime candidates were 1K and 4K size tables. For hardware implementations where memory space is limited, using 1K lookup tables is desirable, but that limits precision to 10 bits. For example, FPGAs tend to have limited memory space (~10 Mbits) compared to PCs (~100 Gbits), and some FPGAs allocate that space in 1K blocks of a number of bits per entry (e.g., 1Kx18 bits). Others [6] have suggested 6 bit precision is sufficient to attain correction performance within 0.1 dB of the Shannon limit. As we can see from comparing the results of Table 5 against Table 3, 10-bit precision doesn't correct as well as double precision, especially for the higher 5/6 rate matrix. However, 10-bit precision does as well as the double precision version for the lower rate (2/3 and 3/5) matrices. Increasing the size of the lookup tables to 4K, resulting in 12-bit precision, generally improves the high rate matrices results, especially for the ETSI matrices. This can be seen from Table 4. Our assumption is that these are cases where the tables couldn't differentiate between a number of adjacent entries due to precision limitations (i.e., when the belief values were low). Furthermore, using 12-bit precision helps the correction algorithm to converge faster, as can be seen in Tables 8-10. Tables 8-10 are grouped by matrix rate and show two values in each table entry. The first value is the number of failures (shown in Tables 3-7) and the second entry is the average number of iterations required for the algorithm to converge. Using 11-bit precision (2K tables) does better than 10-bit precision but not as well as 12-bit precision. Increasing precision, beyond 10-bits, does not have any significant effect on the correction performance for the lower rate (2/3 and 3/5) matrices.

We have explored a number of heuristics in an attempt to increase the performance of these algorithms using these matrices. We found a simple technique that greatly increases the error correction performance for high rate matrices. Simply employ a high estimate of the QBER parameter, which is used to initialize and compute the belief values. For instance, we add 2 or 3 to the QBER, so when the QBER is ~1%, we use an estimate of 4%. We can see the results of this heuristic in Tables 8-10, in the config lines labeled with a "+%" (i.e.,

1K+% and 4K+%). For example, 1K+% indicates 1K lookup tables with an increased QBER estimate. This heuristic allows the use of 1K tables for the ETSI 5/6 rate matrix at 1% QBER and avoids moving to the 3/4 rate matrix for 1% QBER. Using the 3/4 rate matrix yields less secure bits per reconciliation. Using this heuristic for the higher rate matrices also results in slightly faster convergence, thus increased speed performance of the algorithm. Unfortunately using this heuristic with lower rate matrices can give slightly worse performance at the QBER where the matrix begins to fail, but shows no effect in the QBER region where the matrix operates well. For example, the IEEE 11n 2/3 rate matrix at 7% QBER.

Another heuristic we investigated was to monitor the number of checksum errors, cs_err, looking for that number to become small (cs_err < 10) and stop changing for some consecutive iterations (>2). We interpreted that to indicate additional iterations may not be able to achieve convergence due to problems with the belief propagation deciding between only a few bits. Our assumption was that the belief propagation wasn't strong enough to push the values of those few bits over the decision threshold. When this situation occurred, we searched for the smallest, non-negative f_tot values and saved them along with the bit numbers they were associated with. For these small number of checksum errors, there are usually only 2 or 3 possible number of bit errors for each checksum error. For example, 9 checksum errors usually implies either 5 or 7 bit errors. So we first take the 5 lowest, non-negative f_tot values and flip the values of the 5 data bits associated with them and see if that will result in convergence (i.e., **[CS]** = **[DS]**). If that fails, we then do the same for the next 2 lowest, non-negative f_tot values and flip the values of the 2 bits associated with them and see if that will result in convergence. If convergence occurs, we successfully terminate, otherwise we terminate in failure. When this condition occurs, we have seen this heuristic work about 50% to 90% of the time. A further benefit of this heuristic is that it avoids useless iterations that will lead to failure and sometimes it also intercepts instances that will successfully converge in a few more iterations, but converges early with this heuristic. Unfortunately the overall effect is negligible, since this condition occurs infrequently, except for when the QBER is at or near its failure limit and in those cases it tends to work less consistently.

## VI. Summary/Conclusion

We have presented a description of the LDPC forward error correction algorithm adapted for the QKD protocol in a form readily applied by developers. This includes sources for the LDPC matrices as well as lookup tables, data structures and a pseudo code description of the complete LDPC algorithm, both of which are normally absent from LDPC descriptions. We suggest using LDPC matrices that have been defined by the IEEE and ETSI standards organizations for use in various wireless communication standards. We evaluated the QKD error correction performance of these various matrices as a function of the QBER. We provided sufficient detail so that our correction results are reproducible by other researchers. We also discussed the computational precision required for this LPDC algorithm and showed a simple heuristic technique to boost performance with less precision. We have presented complete algorithm descriptions and performance analysis to ease the use of LDPC error correction in future QKD systems. As researchers develop new parity check matrices optimized for QKD that may require different algorithms, similar complete algorithm descriptions and performance analysis will be needed.

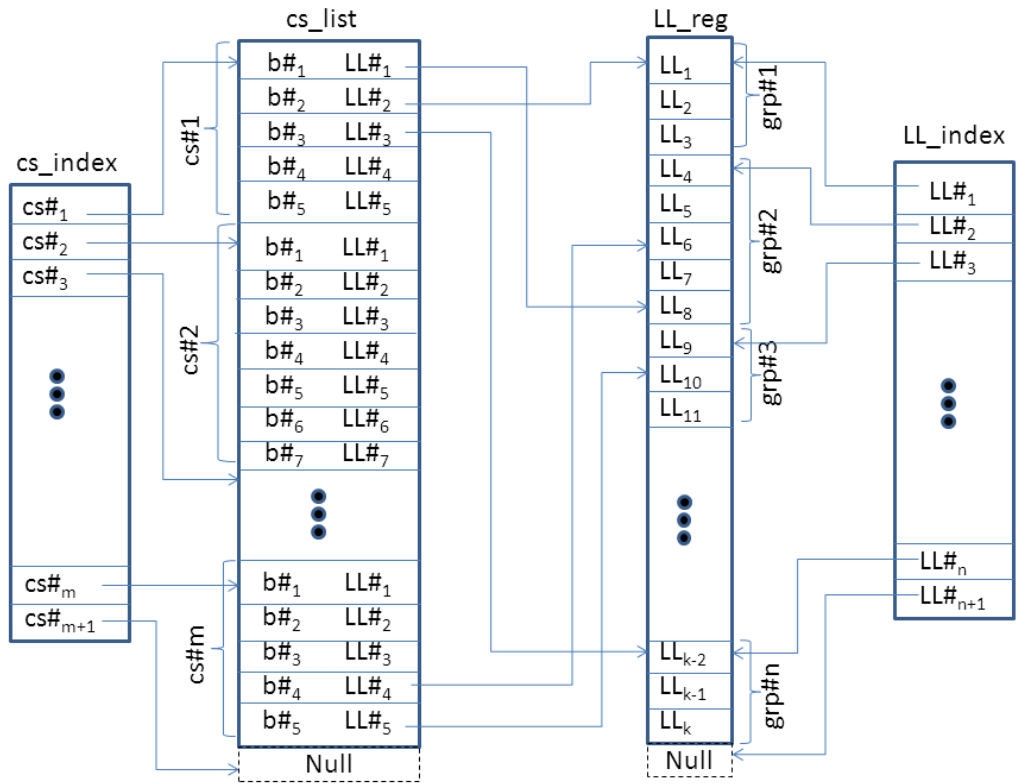

Figure 4. A diagram of the LDPC data mapping lists.

**Table 2.** Matrix characteristics, code rates, number of information bits, number of checksum bits and the expected limits of their correction based on the QBER

| Matrix Source | Code Rate | Info Bits | Number of Checksums | Est QBER Correction |
|---|---|---|---|---|
| IEEE 11n | 5/6 | 1620 | 324 | 2% |
| ETSI DVB | 5/6 | 54000 | 10800 | 2% |
| IEEE 16e | 5/6 | 1920 | 384 | 2% |
| IEEE 11n | 3/4 | 1458 | 486 | 4% |
| ETSI DVB | 3/4 | 48600 | 16200 | 4% |
| IEEE 16e | 3/4 A | 1728 | 576 | 4% |
| IEEE 16e | 3/4 B | 1728 | 576 | 4% |
| IEEE 11n | 2/3 | 1296 | 648 | 8% |
| ETSI DVB | 2/3 | 43200 | 21600 | 8% |
| IEEE 16e | 2/3 A | 1536 | 768 | 8% |
| IEEE 16e | 2/3 B | 1536 | 768 | 8% |
| ETSI DVB | 3/5 | 38880 | 25920 | 10% |

**Table 3.** Double precision baseline failures for 1000 samples (Max 31 iterations for Convergence)

| Matrix Source | Code Rate | 1% | 2% | 3% | 4% | 5% | 6% | 7% | 8% | 9% | 10% | 11% |
|---|---|---|---|---|---|---|---|---|---|---|---|---|
| IEEE 11n | 5/6 | 0 | 413 | 986 | | | | | | | | |
| ETSI DVB | 5/6 | 0 | 0 | 1000 | | | | | | | | |
| IEEE 16e | 5/6 | 0 | 413 | 988 | | | | | | | | |
| IEEE 11n | 3/4 | 0 | 0 | 3 | 206 | 814 | | | | | | |
| ETSI DVB | 3/4 | 0 | 0 | 0 | 0 | 994 | | | | | | |
| IEEE 16e | 3/4 A | 0 | 0 | 10 | 372 | 941 | | | | | | |
| IEEE 16e | 3/4 B | 0 | 0 | 3 | 209 | 852 | | | | | | |
| IEEE 11n | 2/3 | 0 | 0 | 0 | 0 | 10 | 145 | 623 | 944 | | | |
| ETSI DVB | 2/3 | 0 | 0 | 0 | 0 | 0 | 0 | 0 | 993 | | | |
| IEEE 16e | 2/3 A | 0 | 0 | 0 | 0 | 3 | 105 | 543 | 917 | | | |
| IEEE 16e | 2/3 B | 0 | 0 | 0 | 0 | 11 | 163 | 567 | 921 | 997 | | |
| ETSI DVB | 3/5 | 0 | 0 | 0 | 0 | 0 | 0 | 0 | 0 | 0 | 61 | 1000 |

**Table 4.** 12-bit precision, 4k table failures for 1000 samples (Max 31 iterations for Convergence)

| Matrix Source | Code Rate | 1% | 2% | 3% | 4% | 5% | 6% | 7% | 8% | 9% | 10% | 11% |
|---|---|---|---|---|---|---|---|---|---|---|---|---|
| IEEE 11n | 5/6 | 1 | 419 | 988 | | | | | | | | |
| ETSI DVB | 5/6 | 1 | 28 | 1000 | | | | | | | | |
| IEEE 16e | 5/6 | 1 | 417 | 989 | | | | | | | | |
| IEEE 11n | 3/4 | 0 | 0 | 3 | 205 | 815 | | | | | | |
| ETSI DVB | 3/4 | 0 | 0 | 0 | 0 | 996 | | | | | | |
| IEEE 16e | 3/4 A | 0 | 0 | 9 | 365 | 943 | | | | | | |
| IEEE 16e | 3/4 B | 0 | 0 | 2 | 208 | 851 | | | | | | |
| IEEE 11n | 2/3 | 0 | 0 | 0 | 0 | 1 | 10 | 142 | 621 | 944 | | |
| ETSI DVB | 2/3 | 0 | 0 | 0 | 0 | 0 | 0 | 0 | 0 | 1000 | | |
| IEEE 16e | 2/3 A | 0 | 0 | 0 | 0 | 3 | 102 | 543 | 917 | | | |
| IEEE 16e | 2/3 B | 0 | 0 | 0 | 0 | 12 | 159 | 566 | 921 | 997 | | |
| ETSI DVB | 3/5 | 0 | 0 | 0 | 0 | 0 | 0 | 0 | 0 | 0 | 61 | 1000 |

**Table 5.** 10-bit precision, 1k table failures for 1000 samples (Max 31 iterations for Convergence)

| Matrix Source | Code Rate | 1% | 2% | 3% | 4% | 5% | 6% | 7% | 8% | 9% | 10% | 11% |
|---|---|---|---|---|---|---|---|---|---|---|---|---|
| IEEE 11n | 5/6 | 28 | 481 | 991 | | | | | | | | |
| ETSI DVB | 5/6 | 983 | 1000 | 1000 | | | | | | | | |
| IEEE 16e | 5/6 | 15 | 492 | 992 | | | | | | | | |
| IEEE 11n | 3/4 | 0 | 1 | 6 | 205 | 819 | | | | | | |
| ETSI DVB | 3/4 | 26 | 11 | 9 | 10 | 1000 | | | | | | |
| IEEE 16e | 3/4 A | 0 | 0 | 8 | 369 | 943 | | | | | | |
| IEEE 16e | 3/4 B | 1 | 0 | 6 | 258 | 853 | | | | | | |
| IEEE 11n | 2/3 | 0 | 0 | 0 | 0 | 0 | 8 | 137 | 615 | 942 | | |
| ETSI DVB | 2/3 | 0 | 0 | 0 | 0 | 0 | 0 | 0 | 0 | 997 | | |
| IEEE 16e | 2/3 A | 0 | 0 | 0 | 0 | 0 | 2 | 101 | 538 | 916 | | |
| IEEE 16e | 2/3 B | 0 | 0 | 0 | 0 | 0 | 9 | 151 | 556 | 919 | 997 | |
| ETSI DVB | 3/5 | 0 | 0 | 0 | 0 | 0 | 0 | 0 | 0 | 0 | 60 | 1000 |

**Table 6.** 12-bit precision, + extra % QBER, failures for 1000 samples (Max 31 iterations for Convergence)

| Matrix Source | Code Rate | 1% | 2% | 3% | 4% | 5% | 6% | 7% | 8% | 9% | 10% | 11% |
|---|---|---|---|---|---|---|---|---|---|---|---|---|
| IEEE 11n | 5/6 | 0 | 505 | 999 | | | | | | | | |
| ETSI DVB | 5/6 | 0 | 27 | 1000 | | | | | | | | |
| IEEE 16e | 5/6 | 0 | 508 | 998 | | | | | | | | |
| IEEE 11n | 3/4 | 0 | 0 | 2 | 229 | 876 | | | | | | |
| ETSI DVB | 3/4 | 0 | 0 | 0 | 0 | 996 | | | | | | |
| IEEE 16e | 3/4 A | 0 | 0 | 9 | 417 | 969 | | | | | | |
| IEEE 16e | 3/4 B | 0 | 0 | 2 | 250 | 913 | | | | | | |
| IEEE 11n | 2/3 | 0 | 0 | 0 | 0 | 0 | 8 | 156 | 671 | 963 | | |
| ETSI DVB | 2/3 | 0 | 0 | 0 | 0 | 0 | 0 | 0 | 0 | 999 | | |
| IEEE 16e | 2/3 A | 0 | 0 | 0 | 0 | 0 | 2 | 106 | 587 | 952 | | |
| IEEE 16e | 2/3 B | 0 | 0 | 0 | 0 | 10 | 152 | 594 | 952 | 999 | | |
| ETSI DVB | 3/5 | 0 | 0 | 0 | 0 | 0 | 0 | 0 | 0 | 0 | 219 | 1000 |

**Table 7.** 10-bit precision, + extra % QBER, failures for 1000 samples (Max 31 iterations for Convergence)

| Matrix Source | Code Rate | 1% | 2% | 3% | 4% | 5% | 6% | 7% | 8% | 9% | 10% | 11% |
|---|---|---|---|---|---|---|---|---|---|---|---|---|
| IEEE 11n | 5/6 | 1 | 534 | 999 | | | | | | | | |
| ETSI DVB | 5/6 | 2 | 903 | 1000 | | | | | | | | |
| IEEE 16e | 5/6 | 1 | 546 | 998 | | | | | | | | |
| IEEE 11n | 3/4 | 0 | 0 | 2 | 232 | 878 | | | | | | |
| ETSI DVB | 3/4 | 0 | 0 | 0 | 0 | 1000 | | | | | | |
| IEEE 16e | 3/4 A | 0 | 0 | 6 | 418 | 969 | | | | | | |
| IEEE 16e | 3/4 B | 0 | 0 | 2 | 221 | 917 | | | | | | |
| IEEE 11n | 2/3 | 0 | 0 | 0 | 0 | 0 | 8 | 155 | 671 | 963 | | |
| ETSI DVB | 2/3 | 0 | 0 | 0 | 0 | 0 | 0 | 0 | 0 | 0 | 1000 | |
| IEEE 16e | 2/3 A | 0 | 0 | 0 | 0 | 0 | 2 | 109 | 589 | 953 | | |
| IEEE 16e | 2/3 B | 0 | 0 | 0 | 0 | 10 | 147 | 585 | 952 | 999 | | |
| ETSI DVB | 3/5 | 0 | 0 | 0 | 0 | 0 | 0 | 0 | 0 | 0 | 221 | 1000 |

| Matrix Source | Code Rate | Config | QBER 1% | 2% | 3% | 4% | 5% | 6% | 7% | 8% | 9% | 10% | 11% |
|---|---|---|---|---|---|---|---|---|---|---|---|---|---|
| ETSI DVB | 5/6 | Log | 0/6.8 | 0/18.8 | 1000/31.0 | | | | | | | | |
| | | 4k | 1/7.5 | 28/28.1 | 1000/31.0 | | | | | | | | |
| | | 4K+% | 0/7.4 | 27/21.9 | 1000/31.0 | | | | | | | | |
| | | 1K | 983/30.9 | 1000/31.0 | 1000/31.0 | | | | | | | | |
| | | 1K+% | 2/8.5 | 903/30.7 | 1000/31.0 | | | | | | | | |
| IEEE 11n | 5/6 | Log | 0/4.5 | 413/19.3 | 986/30.9 | | | | | | | | |
| | | 4k | 1/4.6 | 419/19.4 | 988/30.9 | | | | | | | | |
| | | 4K+% | 0/5.3 | 505/22.0 | 999/31.0 | | | | | | | | |
| | | 1K | 28/5.8 | 481/21.1 | 991/30.9 | | | | | | | | |
| | | 1K+% | 1/5.5 | 534/22.7 | 999/31.0 | | | | | | | | |
| IEEE 16e | 5/6 | Log | 0/4.6 | 413/19.7 | 988/30.9 | | | | | | | | |
| | | 4k | 1/4.6 | 417/19.8 | 989/30.9 | | | | | | | | |
| | | 4K+% | 0/5.4 | 508/22.4 | 998/31.0 | | | | | | | | |
| | | 1K | 15/5.5 | 492/21.8 | 992/30.9 | | | | | | | | |
| | | 1K+% | 1/5.6 | 546/23.3 | 998/31.0 | | | | | | | | |

Table 8. Performance for various configurations of 5/6 rate matrix; failures/average iterations to converge for 1000 samples (Max 31 iterations for Convergence)

| Matrix Source | Code Rate | Config | QBER 1% | 2% | 3% | 4% | 5% | 6% | 7% | 8% | 9% | 10% | 11% |
|---|---|---|---|---|---|---|---|---|---|---|---|---|---|
| ETSI DVB | 3/4 | Log | 0/4.6 | 0/6.5 | 0/9.1 | 0/14.6 | 994/30.9 | | | | | | |
| | | 4k | 0/4.7 | 0/6.6 | 0/9.3 | 0/14.9 | 996/31.0 | | | | | | |
| | | 4K+% | 0/4.8 | 0/6.5 | 0/9.2 | 0/15.2 | 999/31.0 | | | | | | |
| | | 1K | 26/6.4 | 11/8.2 | 9/11.3 | 0/18.2 | 1000/31.0 | | | | | | |
| | | 1K+% | 0/5.0 | 0/7.0 | 0/10.0 | 0/16.9 | 1000/31.0 | | | | | | |
| IEEE 11n | 3/4 | Log | 0/3.2 | 0/4.6 | 3/7.1 | 206/15.9 | 814/28.3 | | | | | | |
| | | 4k | 0/3.2 | 0/4.6 | 3/7.1 | 205/15.9 | 815/28.4 | | | | | | |
| | | 4K+% | 0/3.5 | 0/4.7 | 2/7.2 | 229/17.0 | 876/29.5 | | | | | | |
| | | 1K | 0/3.3 | 1/4.8 | 6/7.4 | 205/16.3 | 819/28.5 | | | | | | |
| | | 1K+% | 0/3.5 | 0/4.8 | 2/7.4 | 232/17.3 | 878/29.6 | | | | | | |
| IEEE 16e | 3/4 A | Log | 0/3.2 | 0/4.6 | 10/7.8 | 372/19.4 | 941/30.2 | | | | | | |
| | | 4k | 0/3.2 | 0/4.6 | 9/7.8 | 365/19.4 | 941/30.2 | | | | | | |
| | | 4K+% | 0/3.4 | 0/4.7 | 9/7.9 | 417/20.9 | 969/30.6 | | | | | | |
| | | 1K | 0/3.2 | 0/4.7 | 8/7.9 | 369/19.6 | 943/30.2 | | | | | | |
| | | 1K+% | 0/3.4 | 0/4.8 | 6/8.0 | 418/21.0 | 969/30.7 | | | | | | |
| IEEE 16e | 3/4 B | Log | 0/3.2 | 0/4.6 | 3/7.1 | 209/16.2 | 852/29.0 | | | | | | |
| | | 4k | 0/3.3 | 0/4.7 | 2/7.1 | 208/16.2 | 851/29.0 | | | | | | |
| | | 4K+% | 0/3.3 | 0/4.8 | 2/7.3 | 250/17.5 | 913/30.0 | | | | | | |
| | | 1K | 1/3.4 | 0/4.8 | 6/7.4 | 258/17.7 | 853/29.1 | | | | | | |
| | | 1K+% | 0/3.6 | 0/4.9 | 2/7.4 | 221/16.6 | 917/30.0 | | | | | | |

Table 9. Performance for various configurations of 3/4 rate matrix; failures/average iterations to converge for 1000 samples (Max 31 iterations for Convergence)

| Matrix Source | Code Rate | Config | QBER 1% | 2% | 3% | 4% | 5% | 6% | 7% | 8% | 9% | 10% | 11% |
|---|---|---|---|---|---|---|---|---|---|---|---|---|---|
| ETSI DVB | 2/3 | Log | 0/4.0 | 0/4.8 | 0/5.8 | 0/6.9 | 0/8.1 | 0/10.0 | 0/12.9 | 0/19.1 | 993/31.0 | | |
| | | 4k | 0/4.0 | 0/4.9 | 0/5.8 | 0/6.9 | 0/8.3 | 0/10.1 | 0/13.0 | 0/19.3 | 1000/31.0 | | |
| | | 4K+% | 0/4.0 | 0/4.9 | 0/5.8 | 0/6.9 | 0/8.2 | 0/10.1 | 0/13.1 | 0/19.8 | 999/31.0 | | |
| | | 1K | 0/4.4 | 0/5.2 | 0/6.1 | 0/7.3 | 0/8.7 | 0/10.7 | 0/13.8 | 0/20.3 | 997/31.0 | | |
| | | 1K+% | 0/4.0 | 0/5.0 | 0/5.9 | 0/7.1 | 0/8.5 | 0/10.5 | 0/13.7 | 0/20.6 | 1000/31.0 | | |
| IEEE 11n | 2/3 | Log | 0/2.9 | 0/3.5 | 0/4.2 | 0/5.1 | 0/6.5 | 10/8.9 | 145/15.0 | 623/25.3 | 944/30.4 | | |
| | | 4k | 0/2.9 | 0/3.5 | 0/4.3 | 0/5.1 | 1/6.5 | 10/8.9 | 142/15.0 | 621/25.3 | 944/30.3 | | |
| | | 4K+% | 0/2.9 | 0/3.6 | 0/4.3 | 0/5.2 | 0/6.5 | 8/9.0 | 156/15.7 | 671/26.4 | 963/30.6 | | |
| | | 1K | 0/2.9 | 0/3.6 | 0/4.3 | 0/5.2 | 0/6.6 | 8/8.9 | 137/15.1 | 615/25.3 | 942/30.3 | | |
| | | 1K+% | 0/2.9 | 0/3.5 | 0/4.3 | 0/5.2 | 0/6.5 | 8/9.0 | 155/15.8 | 671/26.5 | 963/30.6 | | |
| IEEE 16e | 2/3 A | Log | 0/2.8 | 0/3.6 | 0/4.3 | 0/5.2 | 0/6.5 | 3/8.8 | 105/14.6 | 543/24.4 | 917/30.2 | | |
| | | 4k | 0/2.8 | 0/3.6 | 0/4.3 | 0/5.2 | 0/6.5 | 3/8.8 | 102/14.6 | 543/24.4 | 917/30.2 | | |
| | | 4K+% | 0/2.9 | 0/3.6 | 0/4.4 | 0/5.3 | 0/6.6 | 2/8.9 | 106/15.0 | 587/25.5 | 952/30.6 | | |
| | | 1K | 0/2.9 | 0/3.6 | 0/4.4 | 0/5.3 | 0/6.6 | 2/8.9 | 101/14.7 | 538/24.5 | 916/30.2 | | |
| | | 1K+% | 0/2.9 | 0/3.6 | 0/4.4 | 0/5.3 | 0/6.6 | 2/8.9 | 109/15.1 | 589/25.6 | 953/30.4 | | |
| IEEE 16e | 2/3 B | Log | 0/2.8 | 0/3.4 | 0/4.3 | 0/5.5 | 11/8.0 | 163/14.4 | 567/24.0 | 921/30.0 | 997/31.0 | | |
| | | 4k | 0/2.8 | 0/3.4 | 0/4.3 | 0/5.5 | 12/8.0 | 159/14.3 | 566/24.0 | 921/30.0 | 997/31.0 | | |
| | | 4K+% | 0/2.9 | 0/3.4 | 0/4.4 | 0/5.5 | 10/7.9 | 152/14.4 | 594/24.8 | 952/30.4 | 999/31.0 | | |
| | | 1K | 0/2.8 | 0/3.4 | 0/4.3 | 0/5.5 | 9/8.0 | 151/14.2 | 556/23.9 | 919/30.0 | 997/31.0 | | |
| | | 1K+% | 0/2.9 | 0/3.4 | 0/4.3 | 0/5.5 | 10/7.9 | 147/14.4 | 585/24.7 | 952/30.4 | 999/31.0 | | |

Table 10. Performance for various configurations of 2/3 rate matrix; failures/average iterations to converge for 1000 samples (Max 31 iterations for Convergence)

```
// top level LDPC algorithm
ldpc_algo() {
    LL_init();// initialization
    success = 0;  i = 0;
    for  i < max_loops
    {  cs_msgs_2_bits(); // compute & update belief by chksums
       bit_msgs_2_cs();  // compute & update belief by bits & Bob's bits
       success = converged(); // recomputed chksums & test for convergence
       i++;
       if (success = 1) { print("LDPC converged in %d loops\n", i);  break; }
    }
    if (success = 0)   print("LDPC failed\n");
    return (success);
} // end prog
-----------------------------------------------------------------
// initialize log belief with p=QBER
LL_init() {
     f_init = f_init_list[QBER]; //  Nf*ln( (1-p)/p ),  initial f - belief
     a_init = a_init_list[QBER]; //  Na*ln( (1-2*p) ),  initial a - belief
     for (i=1; i<=k; i++)  // initial all belief regs to channel belief
     LL_reg[i] = a_init; //
} // end fct
-----------------------------------------------------------------
// compute new belief ratios by checksum groups
cs_msgs_2_bits() {
     for (i=1; i<=m; i++) // for each chksum
     {     sign=d[i]; big_alpha=0;
           j1=cs_index[i]; j2=cs_index[i+1];    // get chksum indices
           for (j=j1; j<j2; j++)    // compute Belief for i-th chksum
           {   a1 = LL_reg[cs_list[j].LL_ptr];
               if (a1<0) { sign=1-sign; big_alpha=big_alpha-a1;}
               else big_alpha=big_alpha+a1;
           }
           for (j=j1; j<j2; j++)       // update each belief contribution
           {   a1 = LL_reg[cs_list[j].LL_ptr];
               if (a1 < 0) { p_sign=1-sign; p_alpha=big_alpha+a1;}
               else { p_sign=sign; p_alpha=big_alpha-a1;}
               if (p_alpha<=0)  p_alpha=1;
               if (p_alpha>Ma) p_alpha=Ma;
               if (p_sign==0)  LL_reg[cs_list[j].LL_ptr] = p_alpha;
               else             LL_reg[cs_list[j].LL_ptr] = - p_alpha;
           }
     }   // end loop on checksums
} // end fct
```

```
-------------------------------------------------------------------
// compute new belief values by bit groups
bit_msgs_2_cs() {
    for (i=0; i<n; i++)  // for each info bit
    {  j1= LL_index[i];  j2= LL_index[i+1]; // indices for i-th bit LL_reg group
       f_tot = f_init;                      // init belief value for i-th bit
       for (j=j1; j<j2; j++)               // sum individual beliefs
       {   u= LL_reg[j];
           if (u>0) u = a2f[u];        // convert a-to-f
           else      u = -a2f[-u];
           LL_reg[j] = u;
           f_tot = f_tot + u;
       }
       for (j=j1; j<j2; j++)                    // update individual beliefs
       {     k = f_tot – LL_reg[j];
             if (k<0) {p_sign=1; k=-k;}
           else      p_sign=0;
             if (k<1)  k=1; if (k>Mf) k=Mf;
             if (p_sign == 1)  LL_reg[j] = -f2a[k]; // convert f-to-a
             else  LL_reg[j] = f2a[k];
       }
       if (f_tot < 0)  y1[i] = 1 – y[i];    // update Bob's corrected bits
       else            y1[i] =y[i];
    } // end loop on info bit
} // end fct
-------------------------------------------------------------------
// check if LDPC belief propagation algorithm has converged
converged() {
    success = 1; // init to success
    for (i=1; i<=m; i++) // for each chksum
    {    sum = 0;
         j1=cs_index[i]; j2=cs_index[i+1];  // get chksum indices
         for (j=j1; j<j2; j++)         // compute revised chksum
    { sum = sum XOR y1[cs_list[j].bit#]
    }
    if (sum ≠ c[i]) )
       success = 0; // set to failure
  }
    return (success);
} //  end fct
```

Figure 5. Pseudo code for the complete LDPC algorithm.